\documentstyle[epsfig]{mn}

\newif\ifAMStwofonts

\def\dd{{\rm d}}
\def\Deltaa{\Delta_{\rm A}}
\def\note #1]{{\bf #1]}}
\def\Rs{1}

\ifoldfss
  \ifCUPmtlplainloaded \else
    \NewTextAlphabet{textbfit} {cmbxti10} {}
    \NewTextAlphabet{textbfss} {cmssbx10} {}
    \NewMathAlphabet{mathbfit} {cmbxti10} {} 
    \NewMathAlphabet{mathbfss} {cmssbx10} {} 
  \fi
  \ifAMStwofonts
    \ifCUPmtlplainloaded \else
      \NewSymbolFont{upmath} {eurm10}
      \NewSymbolFont{AMSa} {msam10}
      \NewMathSymbol{\upi}     {0}{upmath}{19}
      \NewMathSymbol{\umu}     {0}{upmath}{16}
      \NewMathSymbol{\upartial}{0}{upmath}{40}
      \NewMathSymbol{\leqslant}{3}{AMSa}{36}
      \NewMathSymbol{\geqslant}{3}{AMSa}{3E}

       \let\le=\leqslant
       
    \fi
  \fi
\fi 

\ifnfssone
  \newmathalphabet{\mathit}
  \addtoversion{normal}{\mathit}{cmr}{m}{it}
  \addtoversion{bold}{\mathit}{cmr}{bx}{it}
  \newmathalphabet{\mathbfit} 
  \addtoversion{normal}{\mathbfit}{cmr}{bx}{it}
  \addtoversion{bold}{\mathbfit}{cmr}{bx}{it}
  \newmathalphabet{\mathbfss} 
  \addtoversion{normal}{\mathbfss}{cmss}{bx}{n}
  \addtoversion{bold}{\mathbfss}{cmss}{bx}{n}
  \ifAMStwofonts
    \ifCUPmtlplainloaded \else
      \UseAMStwoboldmath
      \makeatletter
      \new@mathgroup\upmath@group
      \define@mathgroup\mv@normal\upmath@group{eur}{m}{n}
      \define@mathgroup\mv@bold\upmath@group{eur}{b}{n}
      \edef\UPM{\hexnumber\upmath@group}
      \new@mathgroup\amsa@group
      \define@mathgroup\mv@normal\amsa@group{msa}{m}{n}
      \define@mathgroup\mv@bold\amsa@group{msa}{m}{n}
      \edef\AMSa{\hexnumber\amsa@group}
      \makeatother
      \mathchardef\upi="0\UPM19
      \mathchardef\umu="0\UPM16
      \mathchardef\upartial="0\UPM40
      \mathchardef\leqslant="3\AMSa36
      \mathchardef\geqslant="3\AMSa3E

       \let\le=\leqslant

    \fi
  \fi
\fi 

\ifnfsstwo
  \DeclareMathAlphabet{\mathbfit}{OT1}{cmr}{bx}{it}
  \SetMathAlphabet\mathbfit{bold}{OT1}{cmr}{bx}{it}
  \DeclareMathAlphabet{\mathbfss}{OT1}{cmss}{bx}{n}
  \SetMathAlphabet\mathbfss{bold}{OT1}{cmss}{bx}{n}
  \ifAMStwofonts
    \ifCUPmtlplainloaded \else
      \DeclareSymbolFont{UPM}{U}{eur}{m}{n}
      \SetSymbolFont{UPM}{bold}{U}{eur}{b}{n}
      \DeclareSymbolFont{AMSa}{U}{msa}{m}{n}
      \DeclareMathSymbol{\upi}{0}{UPM}{"19}
      \DeclareMathSymbol{\umu}{0}{UPM}{"16}
      \DeclareMathSymbol{\upartial}{0}{UPM}{"40}
      \DeclareMathSymbol{\leqslant}{3}{AMSa}{"36}
      \DeclareMathSymbol{\geqslant}{3}{AMSa}{"3E}

       \let\le=\leqslant

    \fi
  \fi
\fi 

\ifCUPmtlplainloaded \else
  \ifAMStwofonts \else 
    \def\upi{\pi}
    \def\umu{\mu}
    \def\upartial{\partial}
  \fi
\fi

\title{On the choice of parameters in solar structure inversion}

\author[M.C.~Rabello-Soares et al.]
  {M.C.~Rabello-Soares$^1$,
  Sarbani Basu$^2$, J.~Christensen-Dalsgaard$^{1}$\\
  $^1$Teoretisk Astrofysik Center, Danmarks Grundforskningsfond, and
      Institut for Fysik og Astronomi, Aarhus Universitet, \\
	DK-8000 Aarhus C, Denmark \\
  $^2$Institute for Advanced Study, Olden Lane, Princeton NJ 08540 U.S.A.}

\date{Accepted .
      Received ;
      in original form }

\pagerange{\pageref{firstpage}--\pageref{lastpage}}
\pubyear{1994}

\begin{document}

\maketitle

\label{firstpage}

\begin{abstract}

The observed solar p-mode frequencies provide a powerful
diagnostic of the internal structure of the Sun and permit us to test in
considerable detail the physics used in the theory of stellar structure.
Amongst the most commonly used techniques for inverting 
such helioseismic data
are two implementations of the optimally localized averages (OLA) method,
namely the Subtractive Optimally Localized Averages (SOLA) and Multiplicative
Optimally Localized Averages (MOLA). 
Both are controlled by a number of parameters,
the proper choice of which is very important
for a reliable inference of the solar internal structure.
Here we make a
detailed analysis of the influence of each parameter on the solution and
indicate how to arrive at an optimal set of parameters for a given data set.

\end{abstract}


\section{Introduction}

The observed solar p-mode oscillation frequencies depend on
the structure of the solar interior and atmosphere.
The goal of the inverse analysis is to make inferences about the solar 
structure given these frequencies.
A substantial number of
inversions using a variety of techniques have been reported in the
literature within the last decade
(e.g. Gough \& Kosovichev 1990; 
D\"appen et al. 1991;
Kosovichev 1993;
Dziembowski et al. 1994;
Basu et al. 1997).
Two of the most commonly used inversion methods
are implementations of the optimally localized averages (OLA) method,
originally proposed by Backus \& Gilbert (1968):
the method of Multiplicative Optimally Localized Averages (MOLA),
following the suggestion of Backus \& Gilbert, and
the method of Subtractive Optimally Localized Averages (SOLA),
introduced by Pijpers \& Thompson (1992, 1994).
Both methods depend on a number of parameters
that must be chosen in order to make reliable inferences of
the variation of the internal structure along the solar radius. 
Most authors do  not specify how these
parameters are chosen or how a different choice would affect the solution.
The goal of this work is to make a detailed analysis of the influence of each
parameter on the solution,
as a help towards arriving at an optimal set of parameters
for a given data set.

The adiabatic oscillation frequencies are determined solely by two functions
of position:
these may be chosen as
density $\rho$ and $\Gamma_1 = (\partial \ln p / \partial \ln \rho)_{\rm ad}$
or as any other independent pair of model variables related directly to these
(e.g. Christensen-Dalsgaard \& Berthomieu 1991).
The solar p modes are acoustic waves that propagate in the solar interior
and their frequencies are largely determined by the behaviour of sound speed
$c$. Hence, it is natural to use $c$ as one of the variables,
combined with, e.g., $\rho$ or $\Gamma_1$.
The helium abundance $Y$ is also commonly used,
in combination with $\rho$ or $p/\rho$, $p$ being pressure;
this, however, requires the explicit use of the equation of state,
incomplete knowledge of which could cause systematic errors
(see Basu \& Christensen-Dalsgaard 1997). In this work, we 
consider the inverse problem as defined in terms of
sound speed and density.

\section{Linear Inversion Techniques}

\subsection{The inverse problem}

Inversions for solar structure are based on
linearizing the equations of stellar oscillations
around a known reference model.
The differences in, for example, sound speed $c$ and density $\rho$ 
between the structure of the Sun and the reference
model $(\delta c^2/c^2, \delta \rho/\rho)$ are
then related to the differences
between the frequencies of the Sun and the
model ($\delta\omega_i/\omega_i$) by
\begin{eqnarray}
\frac{\delta\omega_i}{\omega_i} & \! = \! &
\int_0^{\Rs} K_{c^2,\rho}^i (r) \frac{\delta
c^2}{c^2}(r) \dd r 
+ \int_0^{\Rs} K_{\rho,c^2}^i (r) \frac{\delta \rho}{\rho}(r) \dd r \nonumber \\
& + & \frac{F_{\rm surf}(\omega_i)}{Q_i} + \epsilon_i  \; ,
\qquad i = 1, \ldots, M \; ,
\label{eqn:freqdif}
\end{eqnarray}
where $r$ is the distance to the centre, which, for simplicity,
we measure in units of the solar radius $R_{\odot}$.
The index $i$ numbers the multiplets $(n,l)$.
The observational errors are given by $\epsilon_i$,
and are assumed to be independent and Gaussian-distributed
with zero mean and variance $\sigma_i^2$.
The kernels $K_{c^2,\rho}^i$
and  $K_{\rho,c^2}^i$ are known functions of the reference model.
The term in $F_{\rm surf}(\omega_i)$
is the contribution from the uncertainties in the near-surface region
(e.g. Christensen-Dalsgaard \& Berthomieu 1991);
here $Q_i$ is the mode inertia, normalized by the inertia of a
radial mode of the same frequency.

For linear inversion methods, the solution at a given point $r_0$ is
determined by a set of inversion coefficients $c_i(r_0)$, such
that the inferred value of, say, $\delta c^2/c^2$ is
\begin{equation}
\Bigg\langle \frac{\delta c^2}{c^2}(r_0) \Bigg\rangle
=  \sum_i c_i(r_0) \frac{\delta \omega_i}{\omega_i} \; .
\label{eqn:avc2}
\end{equation}
From the corresponding linear combination of 
equations (\ref{eqn:freqdif}) it follows that the 
solution is characterized by {\it the averaging kernel},
obtained as 
\begin{equation}
{\cal K} (r_0,r) = \sum_i c_i(r_0) K_{c^2,\rho}^i(r) \; ,
\label{eqn:avker}
\end{equation}
and also by the cross-term kernel:
\begin{equation}
{\cal C} (r_0,r) = \sum_i c_i(r_0) K_{\rho,c^2}^i(r) \; ,
\label{eqn:crosst}
\end{equation}
which measures the influence of the contribution from
$\delta \rho/\rho$ on the inferred $\delta c^2/c^2$.
The standard deviation of the solution is obtained as
\begin{equation}
\left(
\sum_i c_i^2(r_0) \sigma_i^2
\right)^{1/2} \;.
\label{eqn:error}
\end{equation}
The goal of the analysis is then to suppress the contributions 
from the cross term and the surface term
in the linear combination in equation (\ref{eqn:avc2}),
while limiting the error in the solution.
If this can be achieved
\begin{equation}
\Bigg\langle \frac{\delta c^2}{c^2} (r_0)\Bigg\rangle
\simeq \int_0^1 {\cal K} (r_0,r) \frac{\delta c^2}{c^2}(r) \dd r \; .
\label{eqn:avc2b}
\end{equation}
It is generally required that ${\cal K}(r_0, r)$ has unit integral
with respect to $r$, so that the
inferred value is a proper average of $\delta c^2/c^2$:
we apply this constraint here.
Evidently, the resolution of the inference is controlled by
the extent in $r$ of ${\cal K}$,
the goal being to make it as narrow as possible.

The surface term in equation~(\ref{eqn:freqdif}) may be suppressed by 
assuming that $F_{\rm surf}$ can be expanded in terms of polynomials
$\psi_\lambda$, and constraining
the inversion coefficients to satisfy
\begin{equation}
\sum_i c_i(r_0) Q_i^{-1} \psi_{\lambda}(\omega_i) = 0 \; ,
~~~~~~\lambda=0,1,...,\Lambda 
\label{eqn:surf}
\end{equation}
(D\"appen et al. 1991).
As $F_{\rm surf}$ is assumed to be a slowly varying function
of frequency, we use Legendre polynomials 
of low degree to define the basis functions $\psi_{\lambda}$.
The maximum value of the polynomial degree, $\Lambda$, used in
the expansion is a free parameter of the inversion procedures,
which must be fixed.

There are analogous expressions for the density inversion,
expressing $\langle\delta\rho/\rho (r_0)\rangle$
in terms of the appropriate averaging kernel
obtained as a linear combination of the mode kernels $K_{\rho,c^2}^i$,
and involving a cross term giving the contribution from $\delta c^2/c^2$.
In the case of density inversion,
an additional constraint is obtained by noting that
the mass of the Sun is quite accurately known, the mass
of the reference model being usually fixed at this value;
thus the density difference is generally constrained to satisfy
\begin{equation}
4\pi \int_0^1 \frac{\delta\rho}{\rho}(r) \,\rho(r)\, r^2 \dd r = 0 \; .
\label{eqn:masscon}
\end{equation}
We have found that this constraint is important for stabilizing
the solution.

A number of different inversion techniques can be used for inverting the
constraints given in equation~(\ref{eqn:freqdif}).
We have used two versions of the technique of
Optimally Localized Averages (OLA)
(cf. Backus \& Gilbert 1968)
where the inversion coefficients are determined explicitly.

\subsection {SOLA Technique}

The  aim of the Subtractive Optimally Localized Averages (SOLA)
method (Pijpers \& Thompson 1992, 1994) is to determine the inversion
coefficients so that the
 averaging kernel is an approximation to
a given target ${\cal T}(r_0,r)$, by minimizing
\begin{eqnarray}
\int_0^{\Rs} \left[ {\cal K}(r_0,r) - {\cal T}(r_0,r) \right]^2 \dd r +
\beta \int_0^{\Rs} {\cal C}^2 (r_0,r) \, f(r) \, \dd r \nonumber \\
+ \mu\, {\bar \sigma}^{-2} \sum_{i} c_i^2(r_0) \sigma_i^2 \; ,
\label{eqn:SOLA}
\end{eqnarray}
subject to ${\cal K}$  being unimodular.
Here $f(r)$ is a suitably increasing function of radius aimed at
suppressing the surface structure in the cross-term kernel:
we have used $f(r) = (1 + r)^4$.
Also, $\mu$ is a trade-off parameter, determining the balance between
the demands of a good fit to the target and a small error in the solution;
also, the quantity ${\bar \sigma}^2$ is the average variance,
defined by
\begin{equation}
\bar\sigma^2={\sum_i \sigma_i^2 \over M} \; ,
\label{eqn:avsigma}
\end{equation}
$M$ being the total number of modes.
The second trade-off parameter $\beta$ determines the balance between
the demands of a well-localized averaging kernel and a small cross term.
To suppress the influence of near-surface uncertainties,
i.e., the term in $F_{\rm surf}$,
the coefficients are constrained to satisfy equation~(\ref{eqn:surf}).

We have used target functions defined by
\begin{equation}
{\cal T}(r_0,r) =
A\,  r\, \exp \left[ - \left( \frac{r - r_0}{\Delta(r_0)}
+ \frac{\Delta(r_0)}{2 r_0} \right)^2 \right] \; ,
\label{eqn:target}
\end{equation}
where $A$ is a normalization constant to make the target unimodular.
Thus the target function has its maximum at $r = r_0$ and has almost a
Gaussian shape,
except that it is forced to go to zero at $r = 0$.
The target is characterized by a linear width in the radial
direction: $\Delta(r_0)$ = $\Deltaa c(r_0)/c(r_{\rm A})$,
where $r_{\rm A}$ is a reference radius;
this variation of the width with sound speed
reflects the ability of the modes to resolve solar structure
(e.g. Thompson 1993).  We have taken $r_{\rm A} = 0.2 R_{\odot}$, and
in the following characterize the width by the corresponding
parameter $\Deltaa$.

\subsection {MOLA Technique}

In the case of Multiplicative Optimally Localized Averages (MOLA)
method, the coefficients are found by minimizing
\begin{eqnarray}
\int_0^1 {\cal K}^2 (r_0,r) J(r_0,r) \dd r +
\beta \int_0^1 {\cal C}^2 (r_0,r)  \, f(r) \, \dd r \nonumber \\
+ \mu\, {\bar \sigma}^{-2} \sum_i c_i^2(r_0) \sigma_i^2 \; ,
\label{eqn:MOLA}
\end{eqnarray}
where $J(r_0,r)$ is a weight function 
that is small near $r_0$ and large elsewhere:
\begin{equation}
J(r_0,r) = (r - r_0)^2 \; .
\end{equation}
This, together with the normalization constraint, forces ${\cal K}$ to be
large near $r_0$ and small elsewhere, as desired.
As in equation~(\ref{eqn:SOLA})
$f(r)$ is included to suppress surface structure in the cross-term kernel.
The quantity ${\bar \sigma}^{2}$ is defined by equation~(\ref{eqn:avsigma}).
To suppress the influence of near-surface uncertainties,
i.e., the term in $F_{\rm surf}$,
the coefficients are again constrained to satisfy equation~(\ref{eqn:surf}).
The MOLA technique is generally much more demanding on
computational resources than is the SOLA technique because
it involves analysis of
a kernel matrix which depends on the target $r_0$;
in the SOLA case, the corresponding matrix is independent of $r_0$
and hence need only to be analyzed once, for a given inversion case.

\subsection{Quality measures for the solution}

As seen from the previous sections, the inversions are characterized 
by the free parameters
$\mu$, $\beta$, $\Lambda$ and $\Deltaa$ in the case of SOLA.
These must be chosen
to balance the relative importance of obtaining a well-localized
average of the sound speed (or density) difference,
minimizing the variance of the random error
and reducing the sensitivity of the solution to the second function
(i.e., the cross term) as well as to the surface uncertainties. 

The resolution of the inversion is characterized by the properties of
the averaging kernel (eq.~\ref{eqn:avker}),
which determine the degree to which a well-localized average
of the underlying true solution can be obtained.
Various measures of the width of ${\cal K}$ have been considered
in the literature.
Here we measure resolution in terms of the distance
$\Delta_{\rm qu} = r_{\rm qu}^{(3)} - r_{\rm qu}^{(1)}$
between the upper and lower quartile points of ${\cal K}$;
these are defined such that one quarter of the area under ${\cal K}$
lies to the left of $r_{\rm qu}^{(1)}$ and
one quarter of the area lies to the right of $r_{\rm qu}^{(3)}$.
Furthermore, the location of $r_{\rm qu}^{(1)}$ and $r_{\rm qu}^{(3)}$,
relative to the target location $r_0$, provides a measure of any
possible shift of the solution relative to the target.
For the average of the solution to be well-localized,
it is not enough that $\Delta_{\rm qu}$ be small:
pronounced wings and other structure away from the target
radius will produce nonlocal contributions to the average.
As a measure of such effects in the SOLA case, we consider
\begin{equation}
\chi(r_0) =  \int_0^{\Rs} [ {\cal K}(r_0,r) - {\cal T}(r_0,r)]^2 \dd r \; ,
\end{equation}
which should be small (Pijpers \& Thompson 1994).
In the MOLA case, we introduce
\begin{equation}
\chi'(r_0) =  \int_0^{r_A} {\cal K}^2(r_0,r) \dd r +
\int_{r_B}^1 {\cal K}^2(r_0,r) \dd r \; ,
\end{equation}
where $r_A$ and $r_B$ are defined in such a way that
the averaging kernel has its maximum at $(r_A + r_B)/2$ and
its FWHM is equal to $(r_B - r_A)/2$;
again, a properly localized kernel requires that $\chi'$ is small.

In a similar way, it is useful to define a measure $C(r_0)$ of
the overall effect of the cross term:
\begin{equation}
C(r_0) = \sqrt{ \int_0^{\Rs} {\cal C}^2(r_0,r) \dd r} \; ,
\end{equation}
which should be small in order to reduce the sensitivity of the
solution to the second function.

 \begin{figure}
  \begin{center}
    \leavevmode
  \centerline{\psfig{file=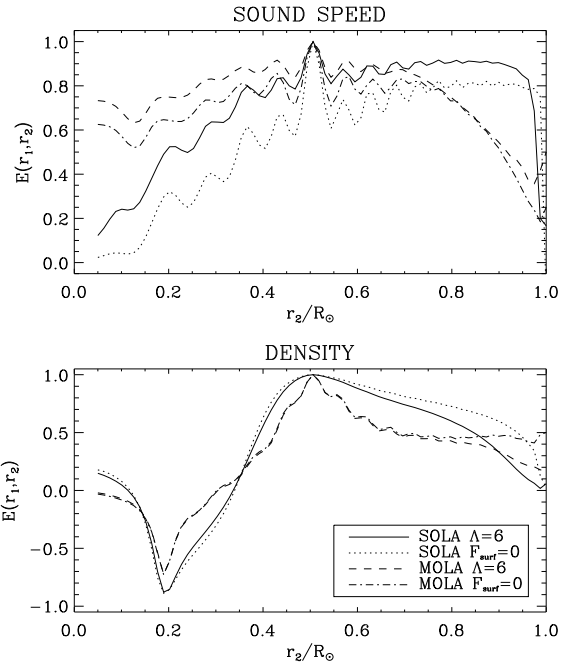,width=8.0cm,height=9.cm}}
  \end{center}
  \caption{\em Error correlation at $r_1 = 0.5 R_{\odot}$
for sound speed and density using SOLA and MOLA,
based on the mode set described in Section~3;
results are shown for a null surface term ($F_{\rm surf} = 0$) and for $\Lambda=6$,
while the remaining inversion parameters have their default values
(cf.\ Section~4).}
  \label{fig:corr}
\end{figure}

It is evident that the overall magnitude of the error in the inferred
solution should be constrained.
However, Howe \& Thompson (1996) pointed out that
it is important to consider also
the {\it correlation} between the errors in the solution
at different target radii.
This arises even if the errors in the original data are uncorrelated:
the errors in the solution at two positions are generally
correlated, because they have been derived from the same set of data.
The normalized correlation function which describes the correlation
between the errors in the solution at $r_1$  and at $r_2$
is defined as:
\begin{equation}
E(r_1,r_2) = \frac{\sum c_i(r_1) c_i(r_2) \sigma_i^2}
{\left[ \sum c_i^2(r_1) \sigma_i^2 \right]^{1/2} \left[ \sum c_i^2(r_2)
\sigma_i^2 \right]^{1/2}} \; .
\end{equation}
Howe \& Thompson showed that correlated errors 
can introduce features into the solution on the scale of the order of 
the correlation-function width. 

Examples of correlation functions are shown in Fig.~\ref{fig:corr}.
For sound-speed inversion, the error correlation generally has a peak
at $r_1 = r_2$ of width corresponding approximately to the width of the
averaging kernel (Fig.~\ref{fig:corr} top).
For density inversion, this peak at $r_1 = r_2$ is much broader than the
averaging-kernel width.
This is a consequence of the difficulty in inferring density
using acoustic-mode frequencies.
There is also a region of strong anti-correlation (Fig.~\ref{fig:corr} bottom). 
This is %
a result of applying the mass-conservation condition (eq.~\ref{eqn:masscon})
since an excess of density in one part of the model has to be 
compensated by a deficiency in another.

\section{Data and models}

The properties of the inversion depends on the mode selection
and errors in the data;
the combination of mode selection and errors is often
described as the {\sl mode set}, in contrast to the
{\sl data set} which in addition contains the data values.
We have based the analysis on the combined LOWL + BiSON mode set
described by Basu et al. (1997). 
Here the modes are in the frequency range 1.5--3.5 mHz,
with degrees between 0 and 99. 
This set in particular provides values for the standard
errors $\sigma_i$ which to a large extent control the
weights given to individual modes;
here $\bar\sigma^2 = 8.6 \times 10^{-11}$ (cf. eq.~\ref{eqn:avsigma}).
In some cases realizations of artificial data were considered;
these were obtained as differences between frequencies of
the proxy and reference models, discussed below,
with the addition of normally distributed random errors
with the variances of the LOWL+BiSON mode set.

Also, but to a far lesser extent, the inversion
depends on the reference model.
We have used Model S of Christensen-Dalsgaard et al. (1996) as our 
reference model.
The model assumes that the Sun has an  age of 4.6 Gyr.
To construct artificial data
for tests of the parameters for solar structure inversion
we adopted as a ``proxy Sun'' another model,
of identical physical assumptions to those in Model S,
but with the lower age of 4.52 Gyr.
With no further modifications,
the ``proxy Sun'' would not include any surface
uncertainties, and hence $F_{\rm surf}$ would have been zero. 
To provide a reasonably realistic model of
this term, the ``proxy Sun'' in addition contained a near-surface
modification based on a simple description of the effects of turbulent
pressure on the frequencies, 
and calibrated to match the actual near-surface contribution
to the difference between the solar frequencies and those of Model S
(cf. Rosenthal 1998).

 \begin{figure}
  \begin{center}
    \leavevmode
  \centerline{\psfig{file=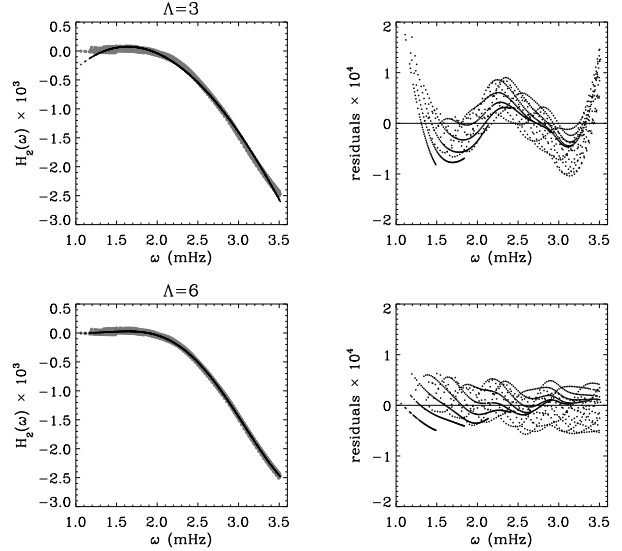,width=8.5cm,height=7.5cm}}
  \end{center}
  \caption{\em On the left-hand side, the grey dots are
$Q_i \delta \omega_i/\omega_i$ after subtracting
the fitted  $H_1(\omega_i/L)$ and the black dots are the fits using $\Lambda=3$
(top) and $6$ (bottom).
On right-hand side, residuals of the $H_2$ fit are plotted.
}
  \label{fig:Gb}
\end{figure}

\section{The choice of inversion parameters}

For a given mode set, the parameters controlling the inversion
must be chosen in a way that, in an appropriate sense,
optimizes the measures of quality introduced in Section~2.4.
Needless to say, this places conflicting demands on the different
parameters, requiring appropriate trade-offs.
Also, it is probably fair to say that no uniquely defined optimum
solution exists.
Here we have chosen what appears to be reasonable parameter sets,
(cf. eqs \ref{eqn:best} and \ref{eqn:bestR}).
The procedure leading to these choices is summarized in Section 4.4;
however, we first justify them by investigating the effect on the 
properties of the inversion of modifications to the parameters
around these values.
This is mostly done in terms of quantities such as error,
error correlation, and kernel properties which do not depend
on the used data values;
however, the effects are also illustrated
by analyses of the artificial data defined in Section~3.

The parameter $\Lambda$ plays a somewhat special role,
in that the suppression of the surface effects is common
to both inversion methods (SOLA and MOLA) and to inversion
for $\delta c^2$ and $\delta \rho$.
For this reason we treat $\Lambda$ separately, in Section~4.1.
The response of the solution to 
the values of the remaining parameters depends somewhat on
the choice of inversion method, and strongly on whether
the inversion is for the sound speed or density difference.
We consider sound-speed inversion in Section~4.2 and density inversion 
in Section~4.3.

\subsection{The choice of $\Lambda$}

Unlike the remaining inversion parameters
the choice of the degree $\Lambda$
used in the suppression of the surface term must 
directly reflect the properties of the data values;
we base the analysis on the near-surface modification introduced
in the artificial data according to the procedure of
Rosenthal (1998) (very similar results are obtained for solar data).
To determine the most appropriate value of $\Lambda$
we consider the frequency-dependent part of the frequency differences.
This is
isolated by noting that according to the asymptotic theory the
frequency differences satisfy 
(e.g.\ Christensen-Dalsgaard, Gough \& Thompson 1989)
\begin{equation}
S_i \frac{\delta \omega_i}{\omega_i} \simeq H_1\left(\frac{\omega_i}{L}\right)
+ H_2(\omega_i) \; ,
\end{equation}
with $L = l + 1/2$, where $l$ is the degree of mode $i$. Here $S_i$
is a scaling factor which in the asymptotic limit is proportional to $Q_i$
and the slowly varying component of
$H_2(\omega_i)$ corresponds to the function $F_{\rm surf}$ in the asymptotic
limit. Thus, by fitting a linear combination of Legendre polynomials to
$H_2$:
\begin{equation}
H_2(\omega_i) 
\sim \sum_{\lambda=0}^{\Lambda} a_{\lambda} P_{\lambda}(\omega_i) \; ,
\end{equation}
we can determine the appropriate value of $\Lambda$ for any given data set.
In practice, we make a non-linear least-squares fit to a sum of two linear
combinations of Legendre polynomials, in $\omega/L$ and
$\omega$ to $S_i \delta\omega_i/\omega_i$,
using a high $\Lambda$ ($\Lambda=16$).
Then we remove $H_1$ from $S_i \delta\omega_i/\omega_i$ and fit now a single
linear combination of Legendre polynomials in $\omega$, looking for
the smallest value of $\Lambda$ that provides a good fit (Fig.~\ref{fig:Gb}).
On this basis, we infer that $\Lambda = 6$ provides an adequate
representation of the surface term; we use this as 
our reference value in the following.
The solar data considered by Basu et al.\ (1997) have a similar behaviour,
and $\Lambda = 6$ is also an appropriate choice in that case.

 \begin{figure}
  \begin{center}
    \leavevmode
  \centerline{\psfig{file=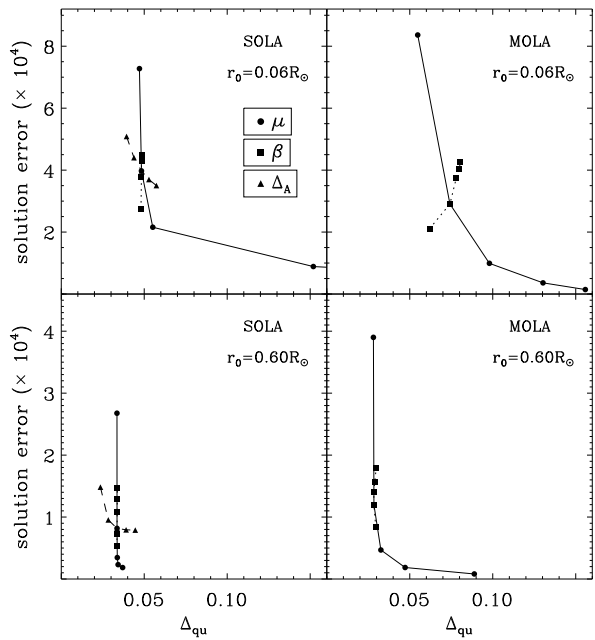,width=9.0cm,height=9.cm}}
  \end{center}
  \caption{\em Trade-off diagram for sound-speed inversion
  at two different radii using SOLA (left) and MOLA (right).
  Values of $\mu$ and $\Deltaa$ increase from top to bottom; and $\beta$ from bottom to top.
In the SOLA case, $\mu$ varies from $10^{-5}$ to $10^{-1}$, $\beta$ from $0.1$ to $1000$
and $\Deltaa$ from $0.04$ to $0.08$. In the MOLA, $\mu$ varies from $10^{-6}$ to $10^{-2}$
and $\beta$ from $0.1$ to $1000$.
Parameters not explicitly mentioned have their reference values
(cf. eq.~\ref{eqn:best}). Note that $\mu=10^{-1}$~in the SOLA
case leads to
a large width when $r_0 = 0.06 R_{\odot}$ and is outside the plot.}
  \label{fig:Cb}
\end{figure}

The constraints imposed by equation~(\ref{eqn:surf}) do not depend 
explicitly on the target location; 
hence it is reasonable that they introduce a contribution to
the errors in the solution that varies little with $r_0$,
leading to an increase in the error correlation.
This is confirmed in the case of sound-speed inversion
by the results shown in the top panel of Fig.~\ref{fig:corr}.
(We note that, in contrast, for density inversion the correlation
{\sl decreases} somewhat with increasing $\Lambda$;
we have no explanation for this curious behaviour, but note that
the density correlation is in any case substantial.)

\subsection{Parameters for sound-speed inversion}

\begin{figure*}
  \begin{center}
    \leavevmode
  \centerline{\psfig{file=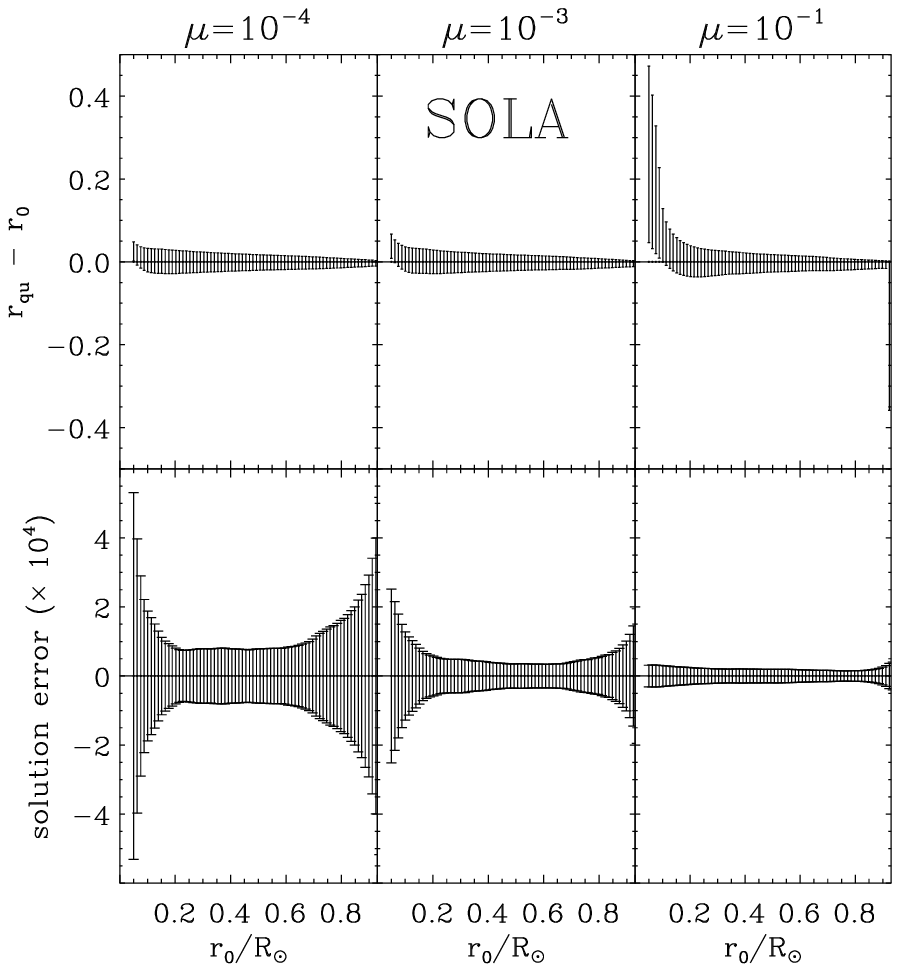,width=8cm,height=8cm}\psfig{file=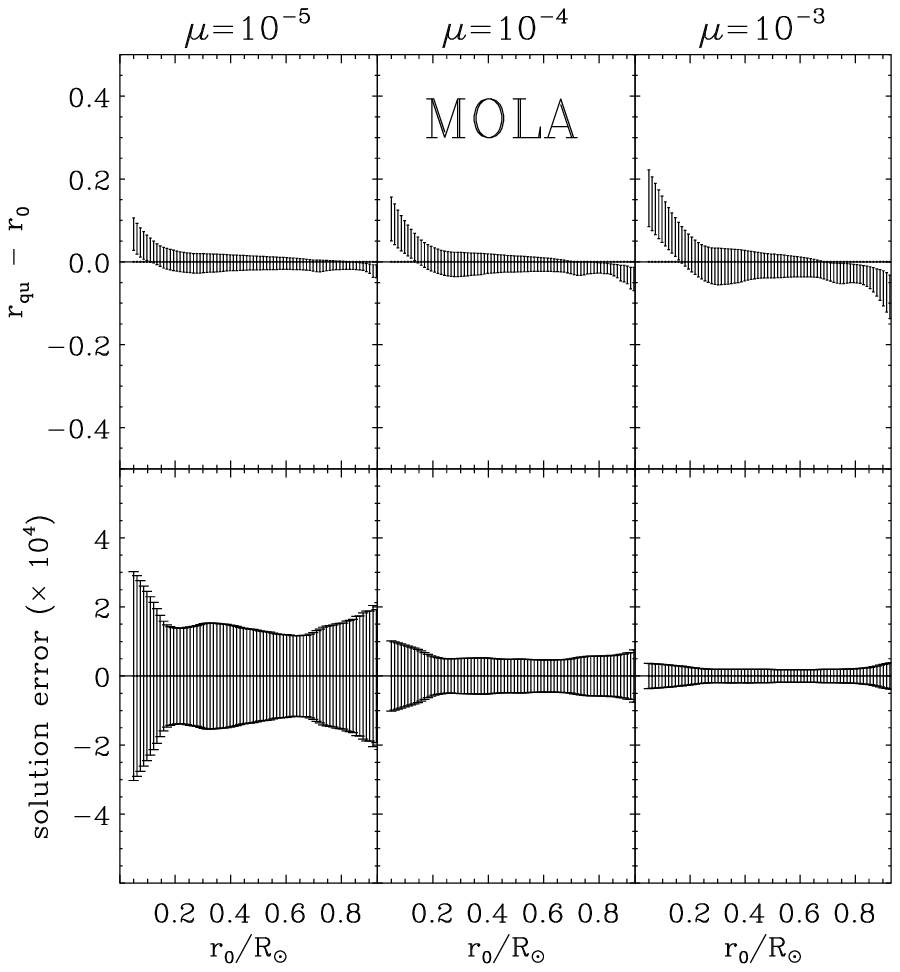,width=8cm,height=8cm}}
  \end{center}
  \caption{\em Resolution of averaging kernels 
  for sound-speed inversion using SOLA (left) and MOLA (right),
  illustrated by the locations 
  $r_{\rm qu}^{(3)} - r_0$ and $r_{\rm qu}^{(1)} - r_0$
  of the upper and lower quartile
  points relative to the target radius (top), and solution error
(bottom), as a function of target radius, for different values of $\mu$.
}
  \label{fig:C2}
 \end{figure*}

\begin{figure*}
  \begin{center}
    \leavevmode
  \centerline{\psfig{file=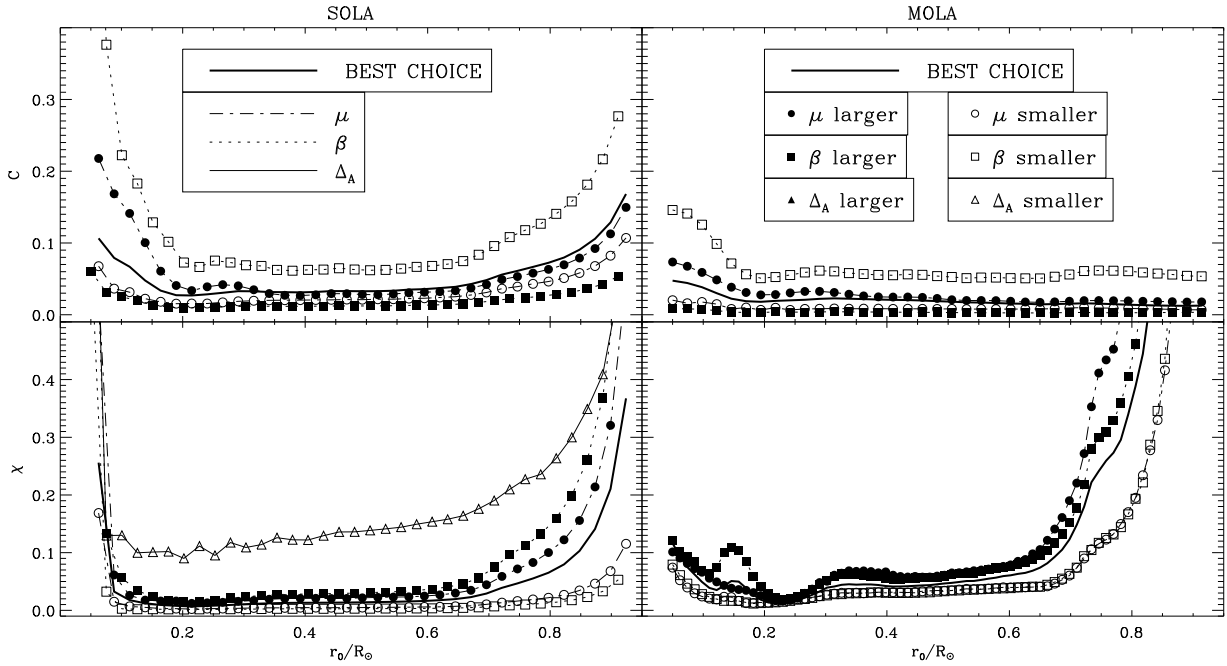,width=17.0cm,height=8.95cm}}
  \end{center}
  \caption{\em Variation of $C$ and $\chi$ ($\chi'$ in the MOLA case),
  for sound-speed inversion.
In the SOLA case, $\mu = 10^{-5}$ and $10^{-3}$, $\beta = 0.1$ and $10$,
and $\Deltaa = 0.04$ and $0.08$; in the MOLA case,
$\mu = 10^{-6}$ and $10^{-4}$ and $\beta = 0.1$ and $10$.
The continuous thick lines use the reference choice of parameters
(cf. eq.~\ref{eqn:best});
parameters not explicitly mentioned have their reference values.
In the SOLA case, a larger value of $\Deltaa$
has no visible effect on  $\chi$, while $C$ is insensitive to 
the change in $\Deltaa$.
}
  \label{fig:Db_Chib}
\end{figure*}

\begin{figure*}
  \begin{center}
    \leavevmode
  \centerline{\psfig{file=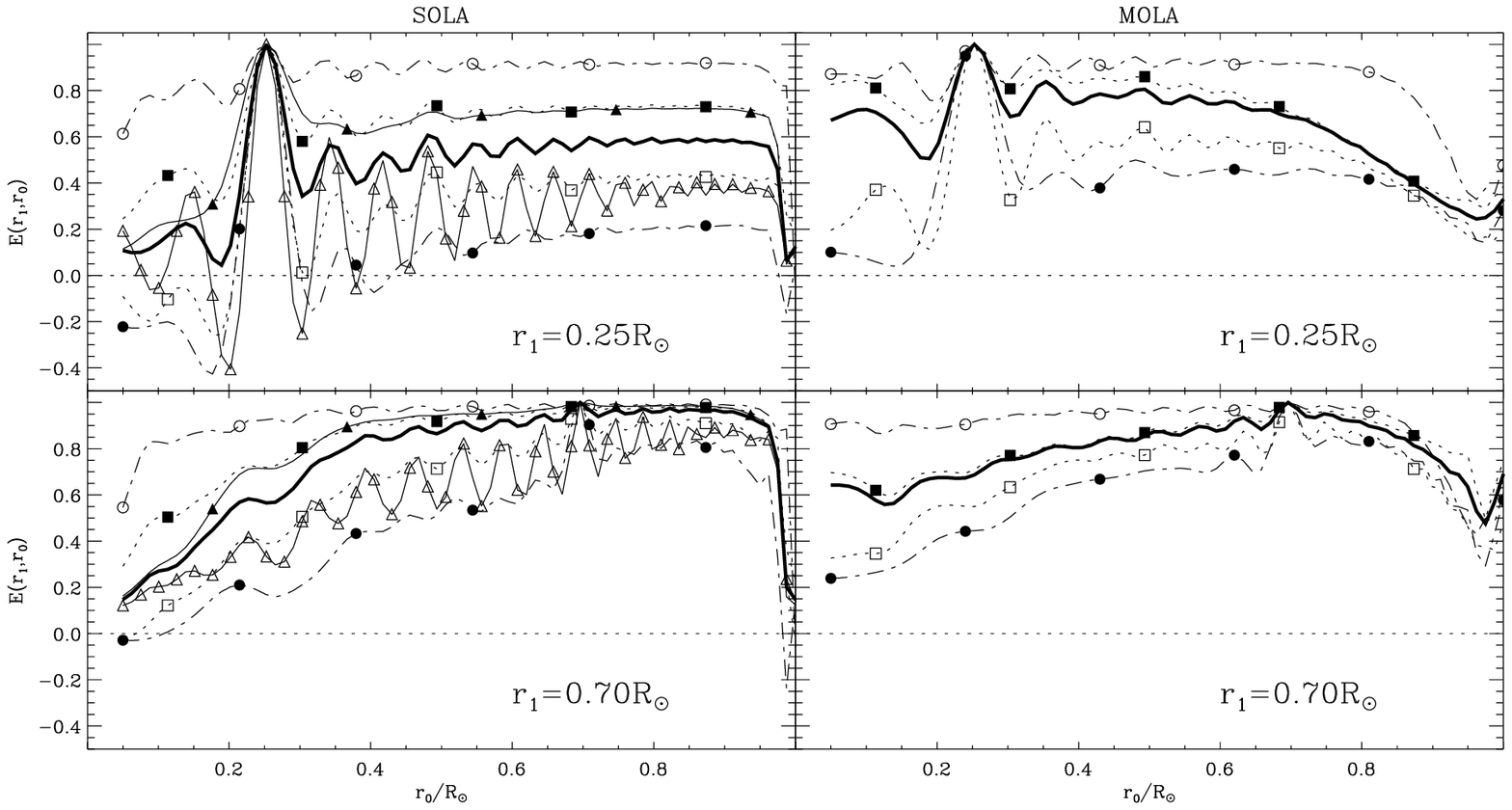,width=17cm,height=8.95cm}}
  \end{center}
  \caption{\em Variation of error correlation with target radius
  using SOLA (left) and MOLA (right) for sound-speed inversion.
In the SOLA case,  $\mu = 10^{-5}$ and $10^{-3}$, $\beta = 0.1$ and $10$,
and $\Deltaa = 0.05$ and $0.07$. In the MOLA case,  $\mu = 10^{-6}$ and
$10^{-4}$ and $\beta = 0.1$ and $10$.
Again, the continuous thick lines
use the `optimal' values (cf. eq.~\ref{eqn:best}),
and parameters not explicitly mentioned have their reference values.
The same symbols are used as in Fig.~\ref{fig:Db_Chib}.
}
  \label{fig:Eb}
  \begin{center}
    \leavevmode
  \centerline{\psfig{file=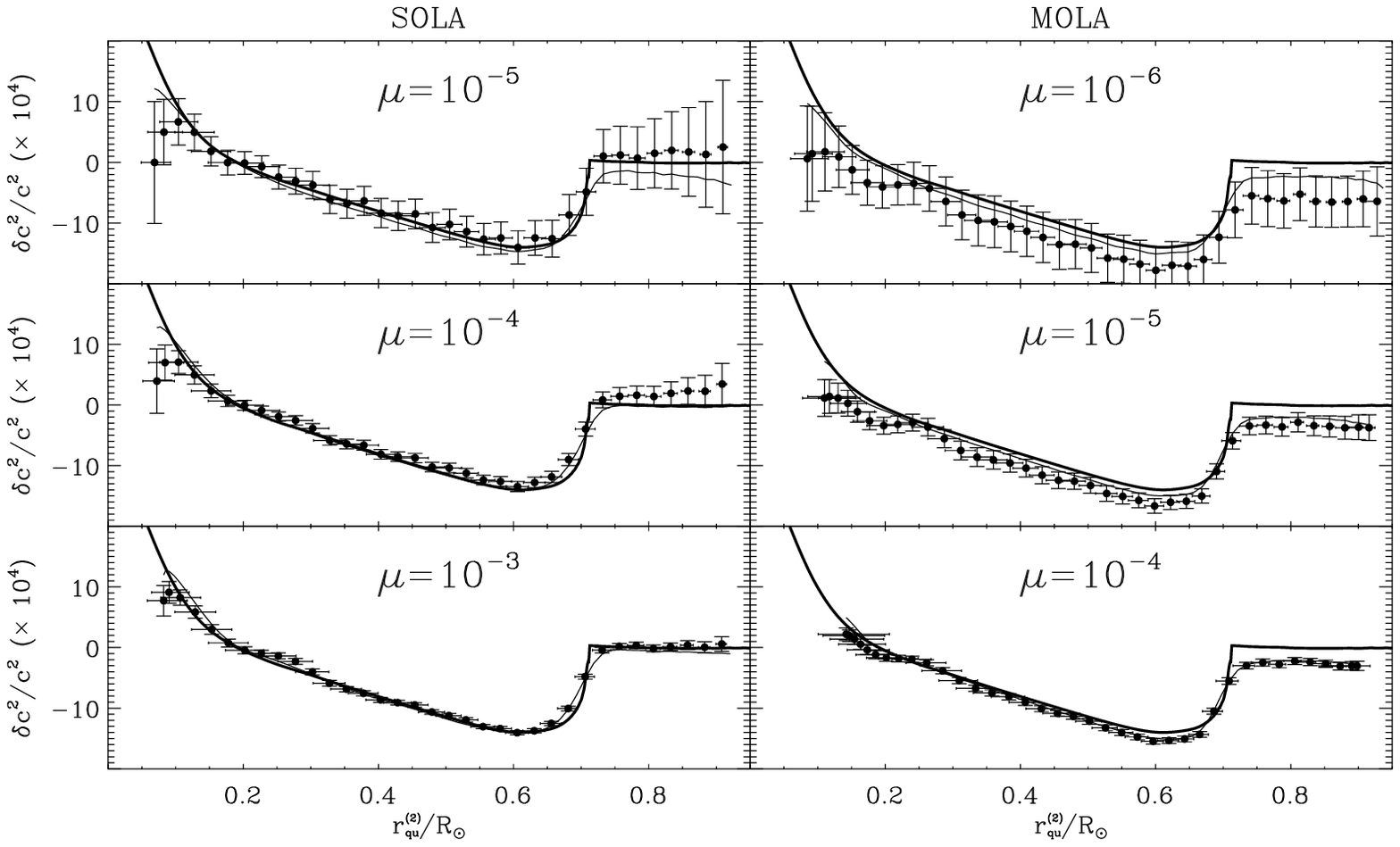,width=17cm,height=10cm}}
  \end{center}
  \caption{\em 
Solution ($\delta c^2/c^2$) versus radius for different $\mu$ using
SOLA (left) and MOLA (right),
for the artificial data described in Section~3 and a single realization
of errors.
Vertical bars show one-sigma errors in the inferred solution,
and are plotted at the second quartile point $r_{\rm qu}^{(2)}$,
while the horizontal bars extend between the first and
third quartile points $r_{\rm qu}^{(1)}$ and $r_{\rm qu}^{(3)}$.
Here $r_{\rm qu}^{(1)}$, $r_{\rm qu}^{(2)}$ and $r_{\rm qu}^{(3)}$
are defined such that 25$\%$, 50$\%$ and 75$\%$ of the
area under the averaging kernel lies to the left of them, respectively.
The middle row uses the `optimal' choice of parameters.
The thick line is the difference
between the theoretical models used. The thin line is the solution of
the inversion without adding errors to the calculated eigenfrequencies.
}
  \label{fig:F2b}
\end{figure*}

\begin{figure*}
  \begin{center}
    \leavevmode
  \centerline{\psfig{file=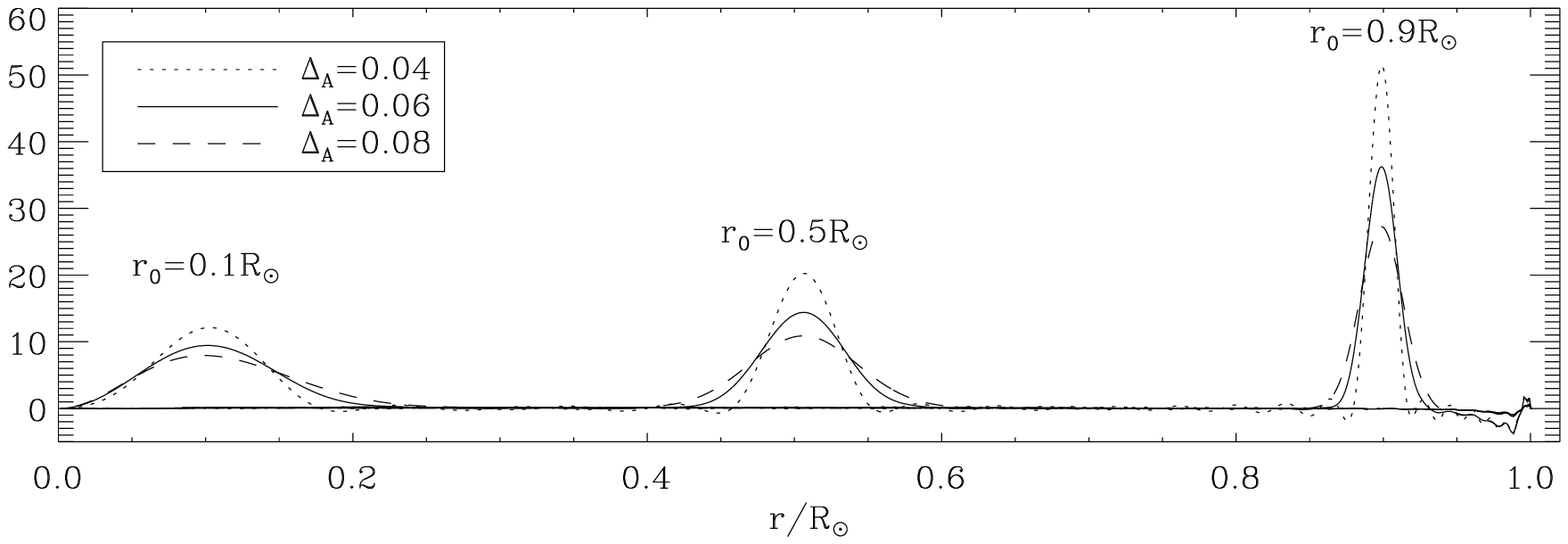,width=17cm,height=6cm}}
  \centerline{\psfig{file=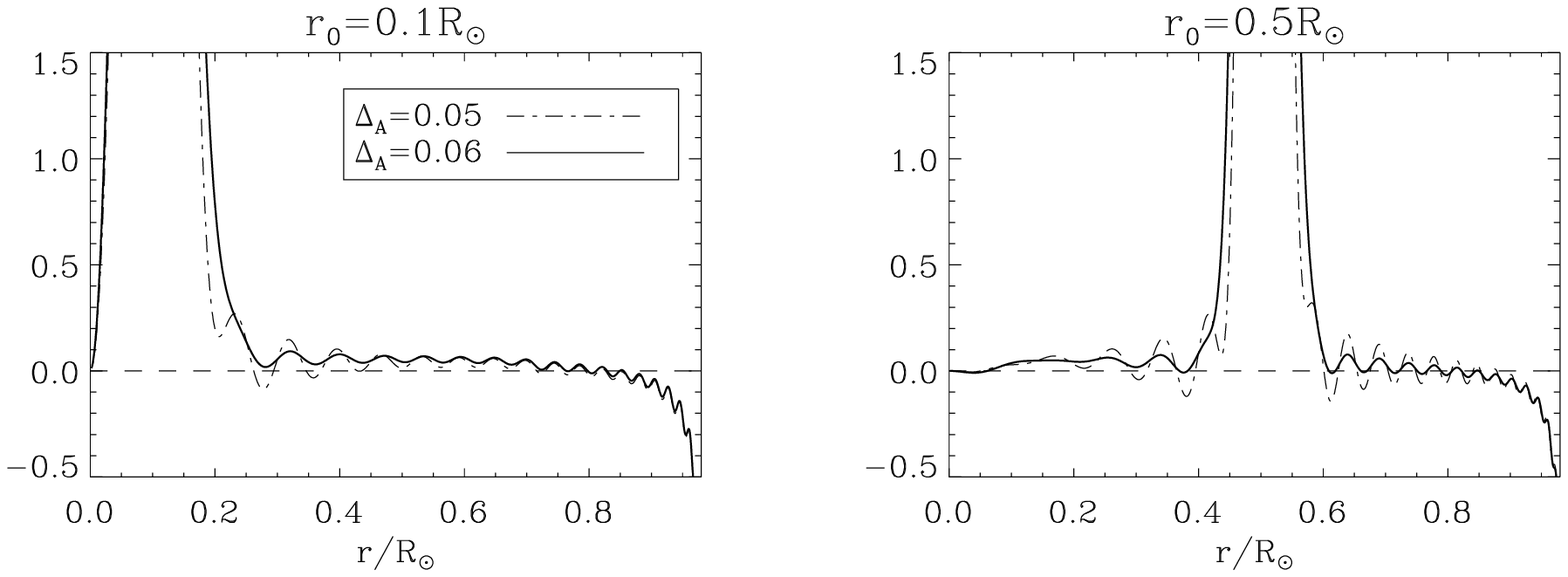,width=17cm,height=6cm}}
  \end{center}
  \caption{\em Top: Averaging kernels for
  SOLA sound-speed inversion at different target radii $r_0$ 
  and different $\Deltaa$.
Note that for a given $\Deltaa$ the width of the
averaging kernel is reduced as the target radius increases. This is
because the width changes with the sound speed (Section~2.2).
Bottom: Detail of the averaging kernels showing their wings
for two target radii and two values of $\Deltaa$.
Near the surface, there is a negative lobe resulting from the fact that
our mode set has modes only for $l \le 99$. 
Closer inspection of the computed kernels
indicates that we do not have sufficient information to go further
than $r_0 \simeq 0.95$.
}
  \label{fig:B2_Bdetb}
\end{figure*}

As reference we use what is subsequently determined to be the
best choice of parameters:
\begin{eqnarray}
\mbox{SOLA}&:&  \Lambda = 6 \, ,\;  \mu = 10^{-4}\, ,\;  \beta = 2 \, ,\; 
 \Deltaa = 0.06 \, ; \nonumber \\
\mbox{MOLA}&:&  \Lambda = 6 \, ,\;  \mu = 10^{-5} \, ,\;  \beta = 1  \; .
\label{eqn:best}
\end{eqnarray}
Effects on the quality measures of varying the parameters 
around these values are illustrated in Figs~\ref{fig:Cb},
\ref{fig:C2}, \ref{fig:Db_Chib} and \ref{fig:Eb};
in addition, Fig.~\ref{fig:F2b} shows results of the analysis
of artificial data (cf. Section~3),
and Fig.~\ref{fig:B2_Bdetb} illustrates properties of selected
averaging kernels.
Throughout, parameters not explicitly mentioned have their reference values.

\subsubsection{Choice of $\mu$}

Generally known as the trade-off parameter, $\mu$
must be determined to ensure a trade-off between the solution error 
(eq.~\ref{eqn:error}) and
resolution of the averaging kernel.
This is typically illustrated in trade-off diagrams such as
Fig.~\ref{fig:Cb}, showing the solution error against resolution
(here defined by the separation $\Delta_{\rm qu}$ between the
quartile points) as $\mu$ varies (circles).
As $\mu$ is reduced, the solution error increases;
the resolution width generally decreases
towards a limiting value which, in the SOLA case, is typically
determined by the target width $\Delta(r_0)$.
On the other hand, for larger values of $\mu$, there is a
strong increase in the width, with a corresponding  very small 
reduction in the solution error.

The behaviour in the trade-off diagram
depends on the target radius $r_0$ considered,
the risk of a misleading solution being particularly serious
in the core or near the surface if $\mu$ is too large.
Thus it is important to look at the trade-off diagram
at different target radii.
This is illustrated in Fig.~\ref{fig:C2},
where solution error (lower panels) and the location of the
quartile points relative to the target radius (upper panels) are plotted.
It is evident that the error increases markedly towards the centre
and surface, particularly in the SOLA case.
In addition, the averaging kernels get relatively broad and there
is a tendency that they are shifted relative to the target location,
particularly near the centre.

Note that the results plotted in Fig.~\ref{fig:C2} use
values of $\mu$ such that  the solution error given by 
SOLA and MOLA techniques are
similar.
Even for small values of $\mu$, the MOLA averaging kernels
do not penetrate as deep into the core as do the SOLA averaging kernels.
The resolution of the averaging kernel is more sensitive to the choice
of $\mu$ using MOLA than using SOLA (cf.\ Fig.~\ref{fig:Cb}).

In addition to the error and resolution, we also need to consider
other properties of the solution.
`Global' properties of the averaging kernels, measured by $\chi$
or $\chi'$ are illustrated in Fig.~\ref{fig:Db_Chib}, together
with the integrated measure $C$ of the cross-talk.
For larger values of $\mu$, these quantities increase,
particularly near the surface.
The strong increase in $\chi'$ (and $\chi$) at large target radii is due 
to the presence of a depression in the averaging kernel near the surface
that increases quickly with $r_0$ 
(cf. Fig.~\ref{fig:B2_Bdetb}), especially for MOLA.
The influence of $\mu$ on the error correlation is illustrated in
Fig.~\ref{fig:Eb}; the correlation evidently increases with
decreasing $\mu$, together with the solution error.
Note that the error correlation increases with the target radius
for any choice of parameters (see also Rabello-Soares, Basu
\& Christensen-Dalsgaard 1998). 
It is slightly smaller using
SOLA than MOLA when their solution errors are similar.

Finally, Fig.~\ref{fig:F2b} shows the inferred solutions for the
sound-speed difference
obtained from the artificial data described in Section~3, for three
values of $\mu$, and compared both with the true $\delta c^2/c^2$ and
the solution inferred for error-free data.
The behaviour of the solution generally reflects
the properties discussed so far.
In particular, it should be noticed that the solution for the
data with errors is shifted systematically relative to the
solution based on the error-free data, reflecting the error correlation,
most clearly visible in the outer parts of the model;
this behaviour illustrates the care required in interpreting 
even large-scale features in the solution at a level comparable
with the inferred errors.
We also note that even the inversion based on error-free data
shows a systematic departure from the true solution, particularly
near the surface.
This appears to be a residual consequence of the imposed near-surface
error, exacerbated by the lack of high-degree modes which might
have constrained the solution in this region.
We have checked this by 
considering in addition artificial data without the imposed near-surface error
(cf.\ Section 3).

\subsubsection {Choice of $\beta$}

The importance of $\beta$ is seen most clearly in Fig.~\ref{fig:Db_Chib},
in terms of the properties of the averaging kernels and the cross term:
as desired, increasing $\beta$ reduces the importance of the cross term
as measured by $C$,
but at the expense of poorer averaging kernels, as reflected
in $\chi$ and $\chi'$.
It should be noticed, however, that
the choice of $\beta$ mainly affects the solution in the core
and near the surface,
while it has little effect in the intermediate parts of the solar interior.
We also note that
the error in the solution  increases with increasing $\beta$
(cf.\ Fig.~\ref{fig:Cb}),
as does the error correlation (see Fig.~\ref{fig:Eb}).
Thus the choice of $\beta$ is determined by the
demand that the cross term be sufficiently strongly suppressed,
without compromising the properties of the averaging kernels and errors.

\subsubsection{Choice of $\Deltaa$}

The SOLA technique has an additional parameter: the width of the target
function at a reference radius (see eq.~\ref{eqn:target}).
The aim of SOLA is to  construct
a well-localized averaging kernel that will provide as good a resolution
as possible. 
As illustrated in Fig.~\ref{fig:Cb}
$\Deltaa$ ensures a trade-off between averaging-kernel resolution
(taking into account also the deviation $\chi$ from the target)
and the solution error.

The effect of $\Deltaa$ on the averaging kernels is
illustrated in Fig.~\ref{fig:B2_Bdetb}.
Evidently, for high $\Deltaa$ the solution is smoothed 
more strongly than at low $\Deltaa$.
However, if $\Deltaa$ is too small, 
the averaging kernel starts to oscillate;
even more problematic is the presence
of an extended tail away from the target radius since it introduces 
a non-zero  contribution from radii far removed from the target.
As in the case of $\mu$, the error increases with increasing resolution
when $\Deltaa$ is reduced. 
On the other hand, 
the error correlation decreases with decreasing $\Deltaa$ 
due to the stronger localization of the solution
(cf.\ Fig.~\ref{fig:Eb}) and develops a tendency to oscillate.

We finally note that $C$ is almost insensitive to $\Deltaa$.

 \begin{figure}
  \begin{center}
    \leavevmode
  \centerline{\psfig{file=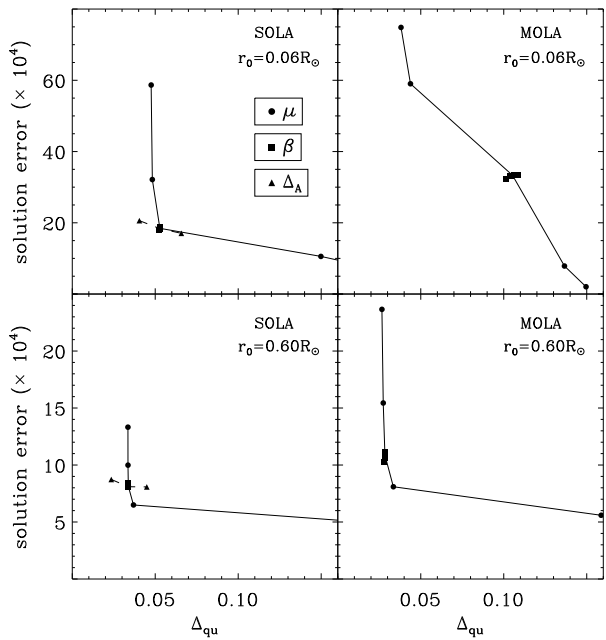,width=9.0cm,height=9.cm}}
  \end{center}
  \caption{\em Trade-off diagram for density inversion
at two different radii using SOLA (left) and 
MOLA (right). Values of $\mu$ and $\Deltaa$ increase from top to bottom:
$\mu$ varies from $10^{-7}$ to $10^{-3}$ (SOLA) 
and from $10^{-9}$ to $10^{-5}$ (MOLA), whereas
$\Deltaa$ varies from $0.04$ to $0.08$. 
Parameters not explicitly indicated correspond to the
best choice (cf. eq.~\ref{eqn:bestR}).
Note that $\mu=10^{-3}$ in the SOLA
case corresponds to a large width, outside the plot.}
  \label{fig:3R}
\end{figure}

\subsection{Density inversion}

\begin{figure*}
  \begin{center}
    \leavevmode
  \centerline{\psfig{file=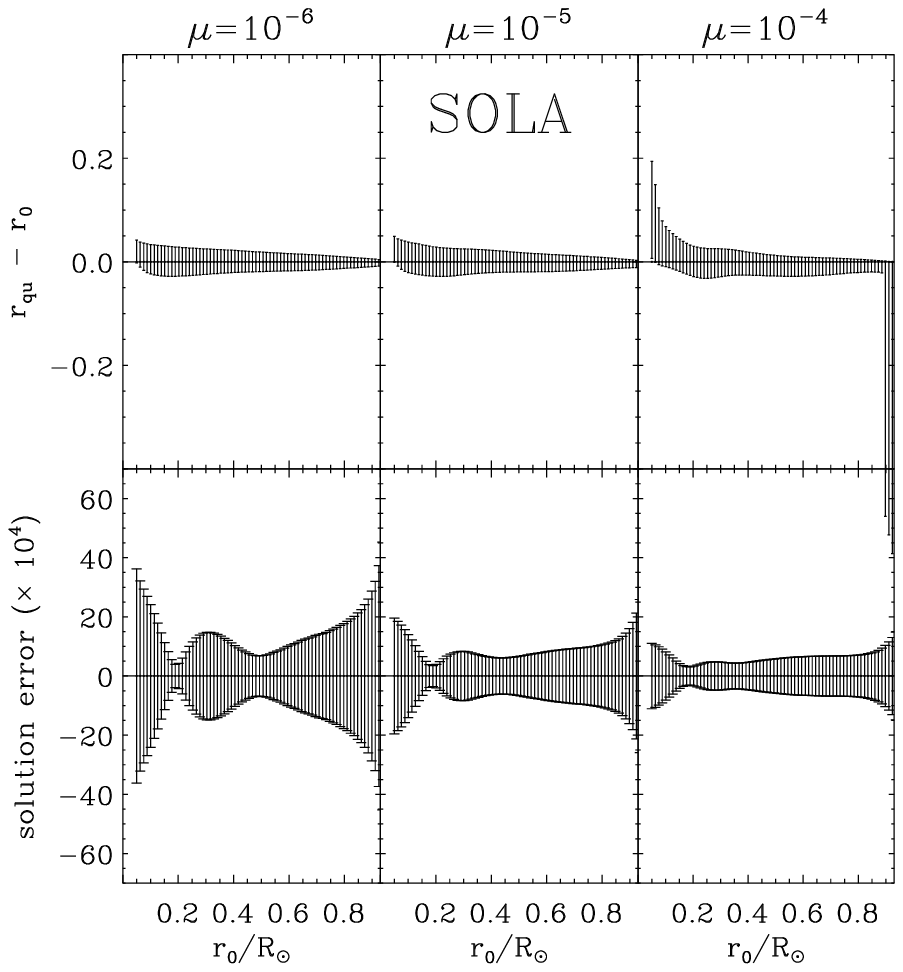,width=8cm,height=8cm}\psfig{file=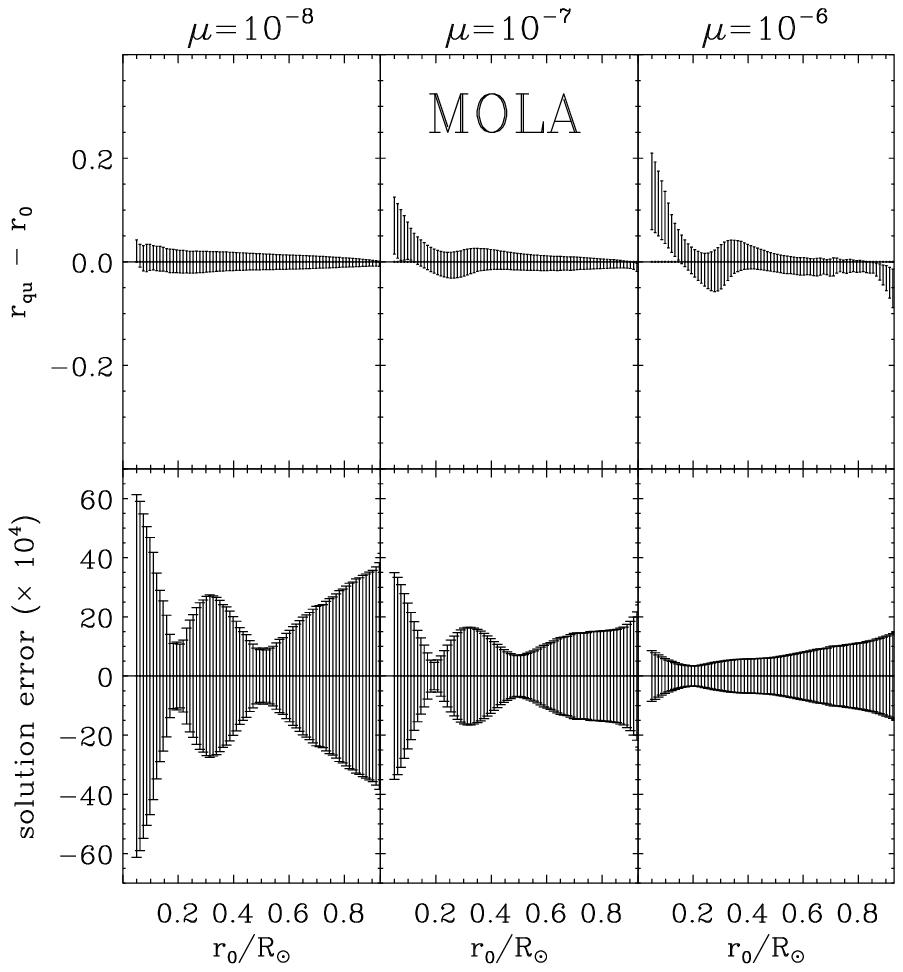,width=8cm,height=8cm}}
  \end{center}
  \caption{\em Resolution of averaging kernels for density inversion
using SOLA (left) and MOLA (right),
  illustrated by the location of the upper and lower quartile
  points relative to the target radius (top), and solution error
(bottom), as a function of target radius, for different values of $\mu$.
}
  \label{fig:4R}
\end{figure*}

\begin{figure*}
  \begin{center}
    \leavevmode
  \centerline{\psfig{file=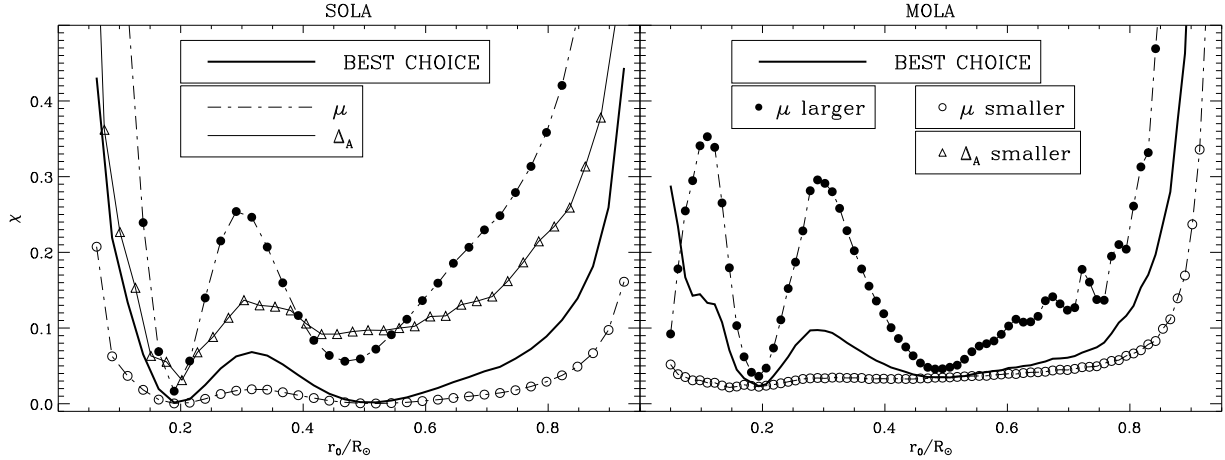,width=17cm,height=6.26cm}}
  \end{center}
  \caption{\em Variation of 
$\chi$ ($\chi'$ in the MOLA case) for density inversion.
In the SOLA case, $\mu = 10^{-6}$ and $10^{-4}$, 
and $\Deltaa = 0.04$; in the MOLA case,
$\mu = 10^{-8}$ and $10^{-6}$. 
The continuous thick line uses the best choice of the parameters
(cf. eq.~\ref{eqn:bestR}).
All the inversions use the best choice of parameters,
except when explicitly indicated.
A larger $\Deltaa$ value cannot be distinguished from the best choice.
}
  \label{fig:5R}
\end{figure*}

\begin{figure*}
  \begin{center}
    \leavevmode
  \centerline{\psfig{file=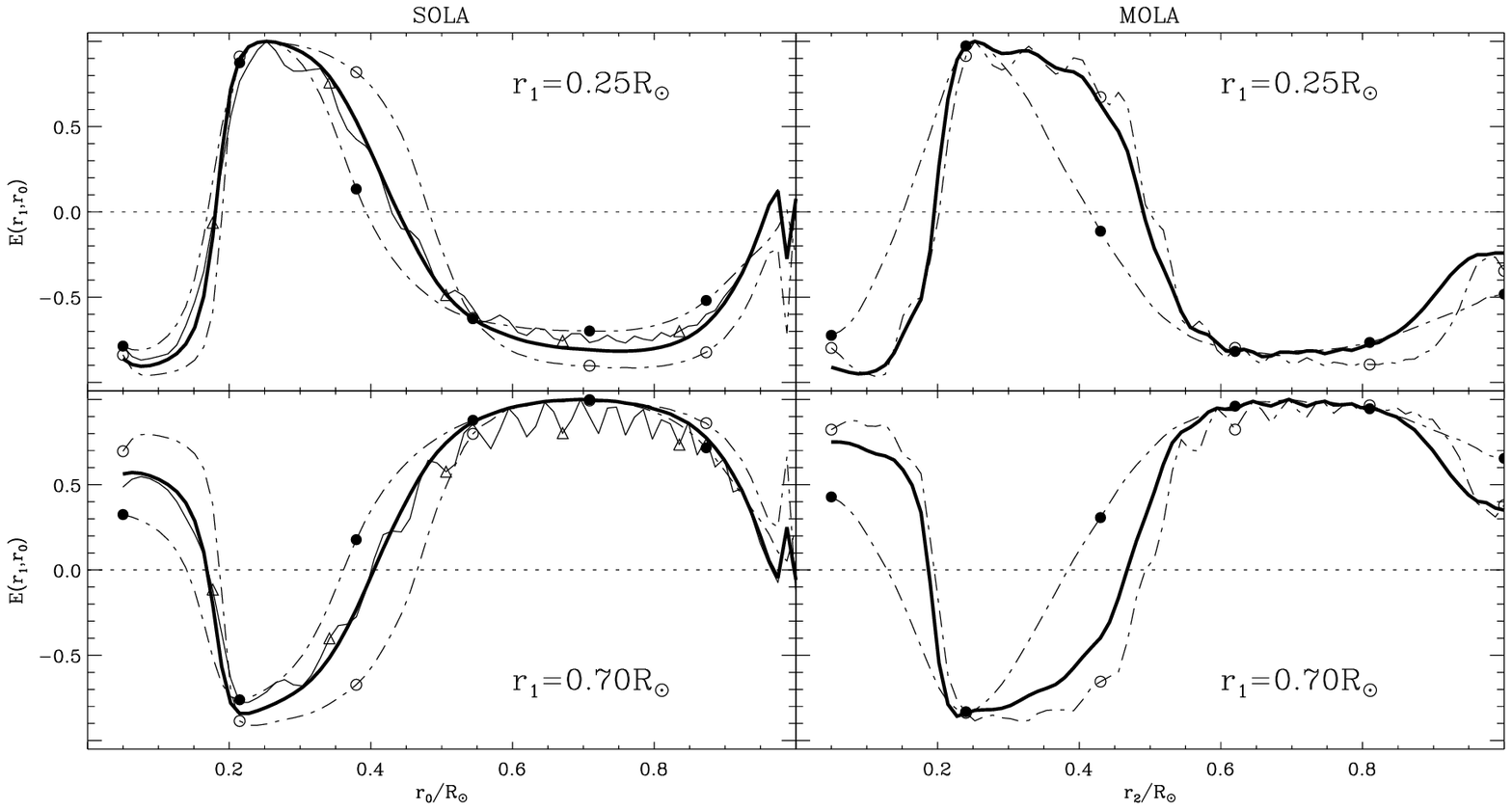,width=17cm,height=8.95cm}}
  \end{center}
  \caption{\em Variation of error correlation with target radius 
using SOLA (left) and MOLA (right) for density inversion.
In the SOLA case,  $\mu = 10^{-6}$ and $10^{-4}$, 
and $\Deltaa = 0.04$. In the MOLA case,  $\mu = 10^{-8}$ and
$10^{-6}$. 
Again, the continuous thick line
uses the `optimal' values (cf. eq.~\ref{eqn:bestR}).
All the inversions use the best choice of parameters,
except when explicitly indicated.
A 
larger $\Deltaa$ value cannot be distinguished from the best choice.
The symbols are as in Fig.~\ref{fig:5R}.
}
  \label{fig:6R}
  \begin{center}
    \leavevmode
  \centerline{\psfig{file=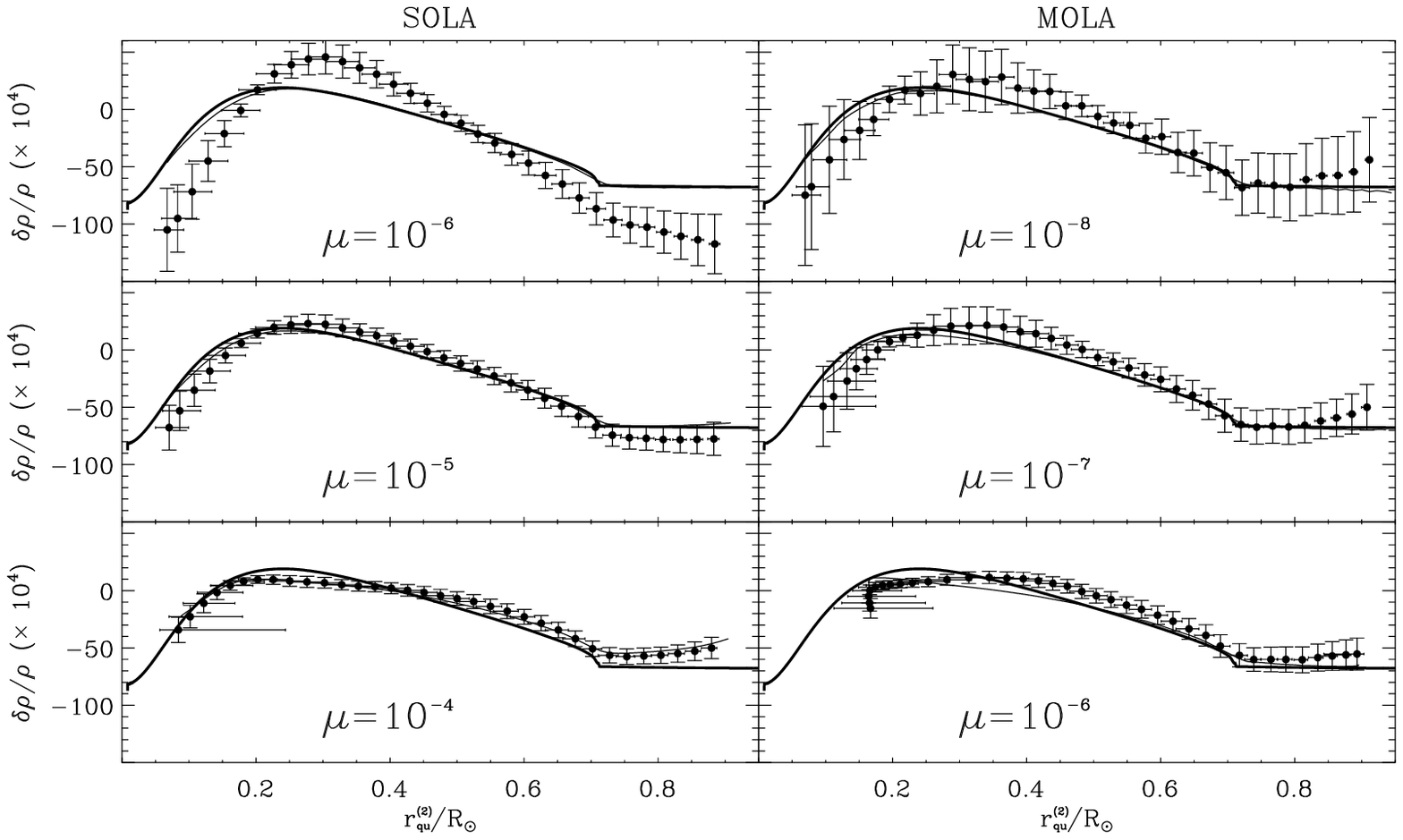,width=17cm,height=10cm}}
  \end{center}
  \caption{\em 
Solution ($\delta \rho/\rho$) versus radius
for different $\mu$ using SOLA (left) and MOLA (right).
The middle row uses the best choice of parameters.
The thick line is the difference
between the theoretical models used. The thin line is the solution of
the inversion without adding noise to the calculated eigenfrequencies.
Note how the error correlation is responsible for a bad solution
when using low $\mu$.
}
  \label{fig:7R}
\end{figure*}

As reference we use what is subsequently determined to be the
best choice of parameters:
\begin{eqnarray}
\mbox{SOLA}&:&  \Lambda = 6 \, ,\;  \mu = 10^{-5} \, ,\;  \beta = 10 \, ,\;  \Deltaa = 0.06 \, ; \nonumber \\
\mbox{MOLA}&:&  \Lambda = 6 \, ,\;  \mu = 10^{-7} \, ,\;  \beta = 50 \; .
\label{eqn:bestR}
\end{eqnarray}
Effects on the quality measures of varying the parameters 
around these values are illustrated in Figs~\ref{fig:3R},
\ref{fig:4R}, \ref{fig:5R} and \ref{fig:6R};
in addition, Fig.~\ref{fig:7R} shows results of the analysis
of artificial data (cf. Section~3),
and Figs~\ref{fig:shoulder} and \ref{fig:8R} illustrate properties of selected
averaging kernels.
Throughout, parameters not explicitly mentioned have their reference values.

In the case of density inversion, we have found that the cross term
is generally small, with little effect on the solution.
In addition, the sound-speed difference between the Sun and
calibrated solar models is typically small in the convection zone
(e.g. Christensen-Dalsgaard \& Berthomieu 1991),
further reducing the effect of the sound-speed contribution in the
density inversion.
As a result,
the effect of the value of $\beta$ on the properties of the inversion
is very modest, although a value of $\beta$ in excess of 1
is required to suppress the remaining effect of the cross term.
Hence, in the following we do not consider the effect of changes to $\beta$,
or the behaviour of $C$.

\subsubsection{Choice of $\mu$}

As for sound-speed inversion, the trade-off parameter $\mu$
must be determined to ensure a trade-off between the solution error and
resolution of the averaging kernel (Fig.~\ref{fig:3R} - circles).
Its behaviour is very similar to that in
the  sound-speed case (Fig.~\ref{fig:Cb}).
As $\mu$ is reduced, the solution error increases
but the resolution width cannot get smaller than a certain value,
which, in the case of SOLA, is the  target width $\Delta(r_0)$.
On the other hand, for larger values of $\mu$, there is a
strong increase in the width, with a corresponding  very small 
reduction in the solution error.

The dependence of the trade-off on target radius $r_0$ is illustrated in
Fig.~\ref{fig:4R}.
If $\mu$ is too large, one may get a misleading solution
especially in the core.
As for sound-speed inversion, the averaging-kernel resolution using MOLA
is more sensitive to variations in $\mu$ than using SOLA.

\begin{figure}
  \begin{center}
    \leavevmode
  \centerline{\psfig{file=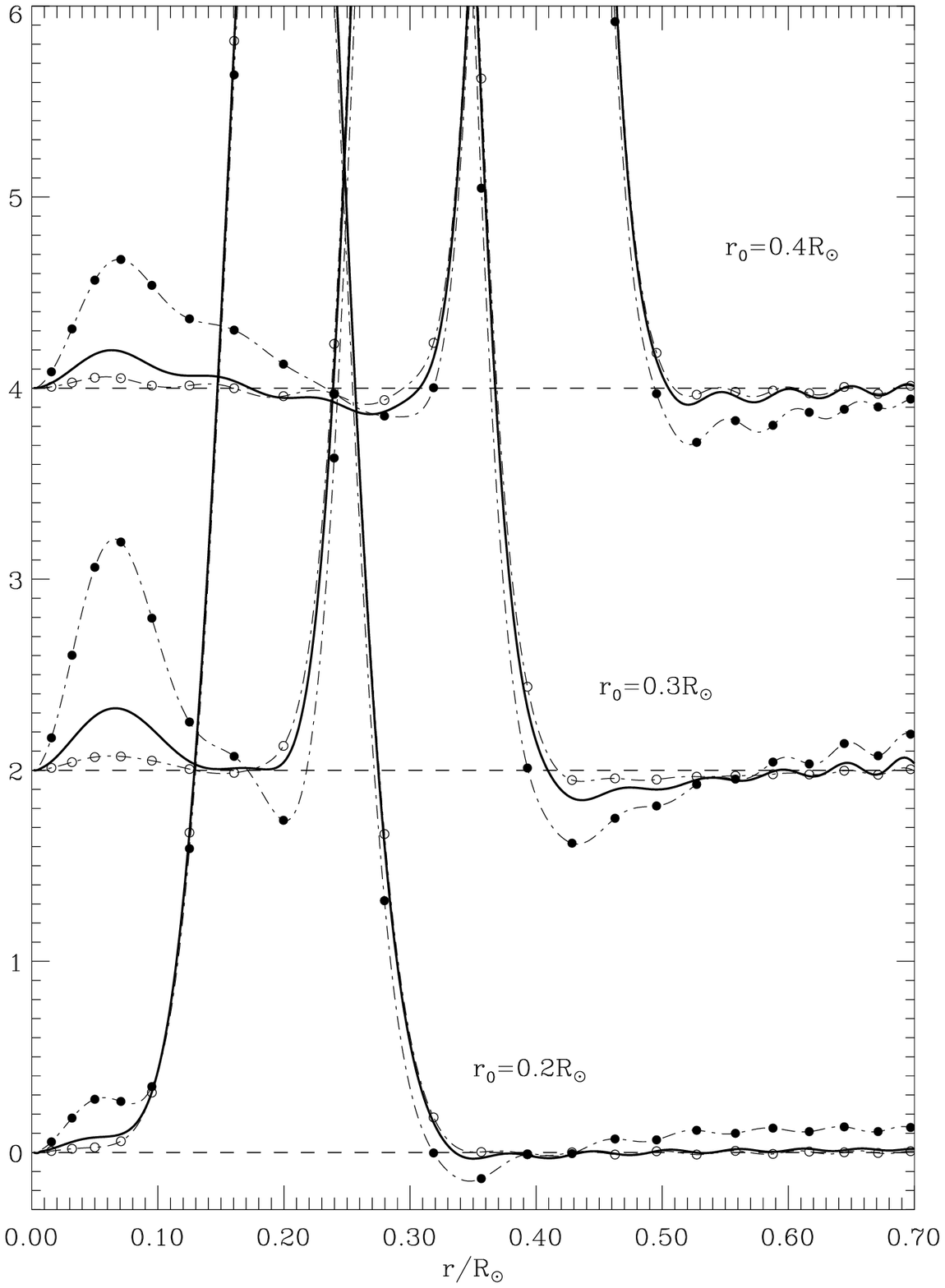,width=8cm,height=10cm}}
  \end{center}
  \caption{\em Averaging kernels for density inversion
  at different target radii $r_0$ and different $\mu$
using the SOLA method.
The averaging kernels for $r_0 = 0.3 R_{\odot}$ and $0.4 R_{\odot}$
are displaced vertically for clarity.
The thick lines use the best choice
of the parameters (eq.~\ref{eqn:bestR}).
The full and empty circles use $\mu=10^{-4}$ and $10^{-6}$,
respectively.
}
  \label{fig:shoulder}
\end{figure}

For larger values of $\mu$, beside the increase in the
averaging-kernel width there is an increase in $\chi$ and $\chi'$ 
(cf.\ Fig.~\ref{fig:5R}).
The bump in $\chi$ and $\chi'$ around $r_0\simeq0.3$ for a large $\mu$
is due to a ``shoulder'' in the averaging kernel
that appears at these target radii, as illustrated in the case of SOLA
in Fig.~\ref{fig:shoulder}.
The error correlation, illustrated in Fig.~\ref{fig:6R},
increases somewhat with increasing $\mu$;
as already noted, the error correlation changes sign 
and is of much larger magnitude for density than for sound speed,
probably as a result of the mass constraint (eq.~\ref{eqn:masscon}).

Examples of inferred solutions, for the
artificial data defined in Section~3, are shown in Fig.~\ref{fig:7R},
together with the true model difference and the difference inferred
from error-free frequency differences.
For a large value of $\mu$ 
the averaging kernel is not well localized and hence affects the solution
(bottom panel);
this is true also for the solution based on error-free data.
For smaller $\mu$ the error clearly increases;
also, particularly in the SOLA case, the effect of the error correlation is
evident.
Thus the solution again deteriorates.
This illustrates how the correlated errors can introduce features into
the solution, showing the importance of limiting the error correlation.

\subsubsection{Choice of $\Deltaa$}

As for the sound-speed inversion, $\Deltaa$ ensures a trade-off between
averaging-kernel resolution and the solution error
in the SOLA technique (cf.\ Fig.~\ref{fig:3R}).
As before, if we choose $\Deltaa$ too small, 
the averaging kernel is poorly localized and it starts to oscillate;
this is reflected in the departure of $\chi$ from the target
(see Fig.~\ref{fig:5R}) and illustrated in more detail,
for selected target radii, in Fig.~\ref{fig:8R}.
On the other hand,
for large $\Deltaa$, the solution is smoothed relative to the one 
for small $\Deltaa$ (due to the low resolution).
The effect on the error correlation of changes in $\Deltaa$ is very modest,
although for small $\Deltaa$ there is a tendency for oscillations
(cf.\ Fig.~\ref{fig:6R}), as was also seen for sound-speed inversion.

\subsection{Summary of the procedure}

\begin{figure*}
  \begin{center}
    \leavevmode
  \centerline{\psfig{file=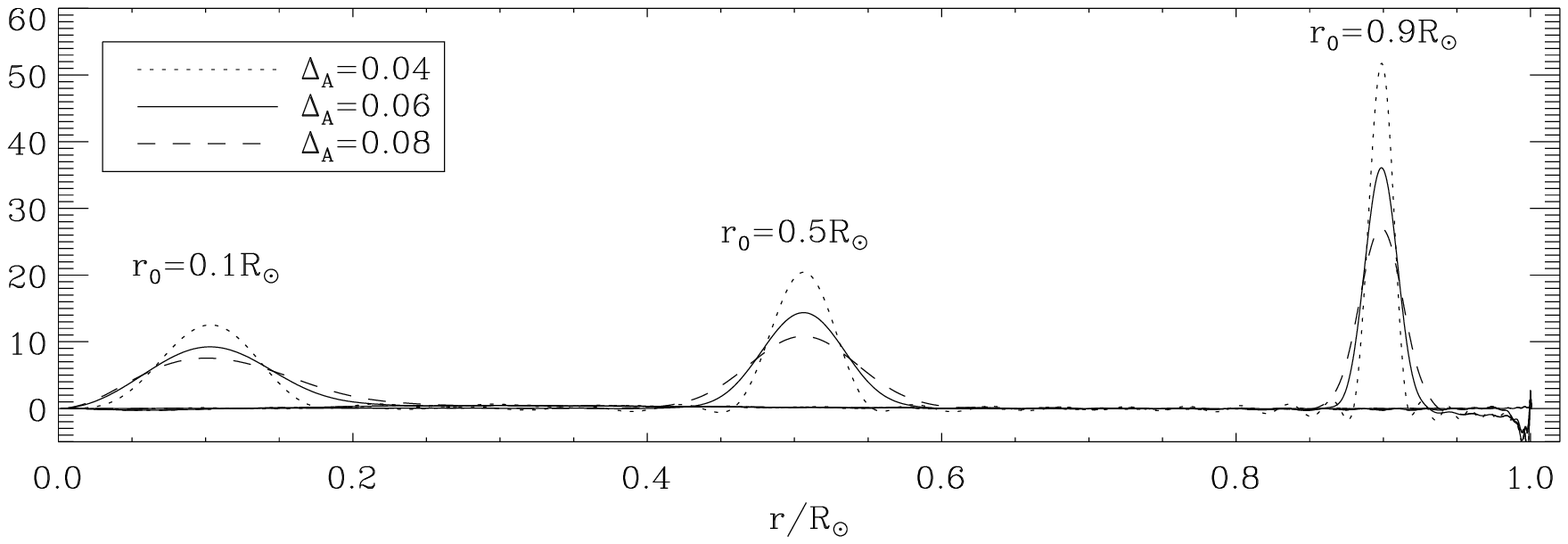,width=17cm,height=6cm}}
  \centerline{\psfig{file=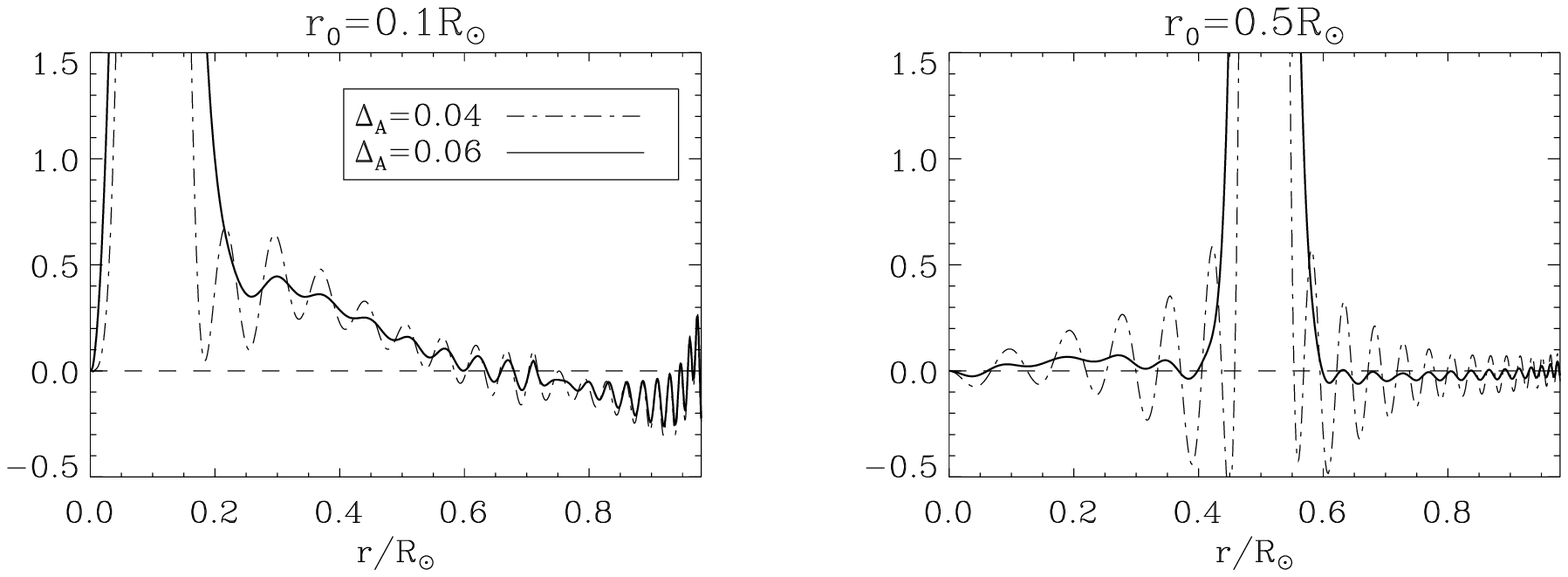,width=17cm,height=6cm}}
  \end{center}
  \caption{\em Top: Averaging kernels for SOLA density inversion
  at different target radii $r_0$ and different $\Deltaa$.
Note that for a given $\Deltaa$ the width of the
averaging kernel is reduced as the target radius increases. This is
because the width changes with the sound speed (see Section~2.2).
Bottom: Detail of the averaging kernels showing their wings
for two target radii and two values of $\Deltaa$.
As in the case of the sound-speed inversion,
the restriction of our mode set to $l \le 99$
causes a negative lobe near the surface.
Closer inspection indicates
that we do not have sufficient information to go further
than $r_0 \simeq 0.92$.
}
  \label{fig:8R}
\end{figure*}

As a convenience to the reader, we briefly summarize the sequence 
of steps which we have found to provide a reasonable
determination of the trade-off parameters for the SOLA and MOLA
methods for inversion for the corrections $\delta c^2$ and $\delta \rho$
to squared sound speed and density:

\begin{itemize}

\item The parameter 
$\Lambda$ is common to both inversion methods and to inversion for
$\delta c^2$ and $\delta \rho$.
Unlike the other parameters, it must directly reflect the properties of
the data values, in terms of the ability to represent the 
surface term (cf. Section 4.1).
It is also largely independent of the choice of
the other parameters: $\mu$, $\beta$ and $\Deltaa$;
thus its determination is a natural first step.
It should be noted, however, that the choice of $\Lambda$ has
some effect on the error correlation (cf. Fig.~\ref{fig:corr}),
generally requiring that $\Lambda$ be kept as small as possible.

\item
The second step is the determination of $\mu$,
whose value is the most critical to achieve a good solution.
As described in Sections 4.2.1 and 4.3.1, it
must be determined to ensure a trade-off between the solution error and
resolution of the averaging kernel 
(Figs~\ref{fig:Cb} and~\ref{fig:3R} - circles)
at representative target radii $r_0$
(see also Figs~\ref{fig:C2} and~\ref{fig:4R}).
In addition, we need to consider the broader properties of the
averaging kernels as characterized by
$\chi$ (SOLA) or $\chi'$ (MOLA) and the cross-talk
quantified by $C$ (Figs~\ref{fig:Db_Chib} and~\ref{fig:5R}),
as well as the error correlation (Figs~\ref{fig:Eb} and~\ref{fig:6R}).

\item 
The next step is to find $\beta$ which is determined by the
demand that the cross term be sufficiently strongly suppressed,
without compromising the properties of the averaging kernels and errors.
Its effect on the properties of density inversion is very modest.

\item
Finally, the SOLA technique has an additional parameter: $\Deltaa$,
which ensures a trade-off between averaging-kernel resolution
and the solution error. $\Deltaa$ is typically decreased until
the averaging kernels are poorly localized and start to
oscillate (Figs~\ref{fig:B2_Bdetb} and~\ref{fig:8R}) which
is reflected in the departure of $\chi$ from the target.

\end{itemize}

After this first determination of $\mu$, $\beta$ and possibly $\Deltaa$,
we go back to step number 2,
determining $\mu$ using now the new values of $\beta$ and $\Deltaa$. 
The procedure obviously requires initial values of 
$\beta$ and (for SOLA) $\Deltaa$:
we suggest $\beta=10$ and $\Deltaa=0.06$ or larger.

Although the measures of quality of the inversion are
essentially determined by the mode set (modes and errors),
the determination of the parameters must also be such as to keep 
the solution error sufficiently small 
to see the variations in
the relative sound-speed or density differences, which in
the case of solar data and a suitable reference model
could be as small as $10^{-3}$ and $5\times10^{-2}$ respectively.
Furthermore, to enable comparison of the solution of the inversion 
of two different data sets, the solution errors should be similar.
To obtain an impression of the quality of the solution and
the significance of inferred features,
we also strongly recommend analysis of artificial data
for suitable test models, including comparison 
of the inferred solutions for the selected parameters
with the true difference between the models and with
the solution inferred for error-free data
(see Figs~\ref{fig:F2b} and~\ref{fig:7R}).

\section{Conclusion}

Appropriate choice of the parameters controlling inverse analyses
of solar oscillation frequencies is required if reliable inferences
are to be made of the structure of the solar interior.
This choice must be based on the properties of the solution,
as measured by the variance and correlation of the errors,
by the resolution of the averaging kernels 
and by the influence of the cross-talk.
We have considered a mode set representative of current inverse
analyses and investigated the properties of the inversion,
as well as the solution corresponding to a specific set of artificial data.
By varying the parameters we have
obtained what we regard as a reasonable choice of parameters
(cf.\ eqs~\ref{eqn:best} and~\ref{eqn:bestR});
this was verified by considering in some detail the sensitivity
of the relevant measures of the quality of the inversion to 
changes in the parameters.
The analysis also illustrated that an unfortunate choice of
parameters may result in a misleading inference of the 
solar sound speed or density; furthermore it became evident that
the correlation between the error in the solution at different 
target location plays an important role, even for our optimal
choice of parameters, and hence must be taken into account
in the interpretation of the results (see also Howe \& Thompson 1996).

The meaning of the parameters is evidently closely related to
the precise formulation of the inverse problem. 
For example, it would be possible to introduce weight functions
in the integrals in equations~(\ref{eqn:SOLA}) and (\ref{eqn:MOLA}),
to give greater weight to specific aspects of the solution.
The need for such refinements is suggested by the fact that
the properties of the solution depends rather sensitively on
the target location $r_0$.
More generally, it is likely that the best choice of parameters
may depend on the target location, further complicating the
analysis and (particularly in the SOLA case) increasing the
computational expense.

The procedure adopted here is evidently somewhat ad hoc,
although we have attempted a logical sequence in the 
order in which the parameters were chosen.
A more systematic approach, making use of objective criteria,
would in principle be desirable.
However, even in the considerable simpler case of inversion for
a spherically symmetric rotation profile, characterized essentially
just by the parameters $\mu$ and possibly $\Deltaa$,
such objective determination of the parameters has so far met
with little success in practice
(see, however, Stepanov \& Christensen-Dalsgaard 1996 and Hansen 1996).
On the other hand, it is far from obvious that an objectively
optimal solution to the inverse problem exists, for a given data set:
the best choice of parameters may well depend on the specific aspects
of the solar interior that are being investigated.
It is important, however, that the error and resolution properties
of the solution be kept in mind in the interpretation of the results;
indeed, the immediate availability of
measures of these properties is a major advantage of 
linear inversion techniques such as those discussed here.

\section*{Acknowledgments}

We are very grateful to M. J. Thompson and an anonymous referee
for constructive comments on earlier versions of the manuscript.
This work was supported in part
by the Danish National Research
Foundation through its establishment of the Theoretical Astrophysics Center.

\bsp 

\label{lastpage}

\end{document}